# Semi-massless fermions tunneling through a gate barrier in graphene

**Bumned Soodchomshom**

Thailand Center of Excellence in Physics, Commission Higher on Education, Ministry of Education, Bangkok 10400, Thailand

**Abstract**

In the case of the strongly deformed graphene, gapless graphene may turn to gapped graphene at the critical deformation. We find that, like semi-massless fermions, electrons in the deformed graphene, at the critical point, mimic the dispersions of the massless fermions in one direction and the massive fermions in the other. Our predicted dispersion formula is the generalization of the previously predicted formula. The behavior of the particle-like semi-massless fermions tunneling through a gate barrier is contrasted with that of the (pure) massless fermions tunneling through a gate barrier in the original graphene. This is due to the effect of the combination between the massless and the massive particle dispersions at the critical deformation.

**1. Introduction**

The massless Dirac fermions tunneling in the condensed matter became a topic of interest since graphene, a monolayer of graphite, has been fabricated [1]. Because of the specific structure, the honeycomb lattice, the effective electrons in graphene mimic the 2D massless Dirac fermions [2, 3], leading to the several novel phenomena in the condensed matter physics. Klein effect, tunneling with out back scattering, is one of the behaviors to confirm the existence of the massless relativistic particles in graphene. Recently, much interest has been focused on the electronic properties in the honeycomb lattices created with cold atoms in optical lattices [4-6] and graphene where they are deformed [7-10]. The deformed honeycomb-like lattice and the deformed graphene give rise to the interesting effective band structures. The gapless energy band turning to the gapped energy band at the critical deformation has been predicted [4-10]. Also, Klein tunneling has been investigated in the deformed honeycomb lattice which is possibly created with cold atoms in optical lattices by the three layer beams [4]. The effect of the deformation brings the transport properties depending on the deformed structure. The Klein effect disappears, when energy gap is opened up between the valence and the conduction bands, due to the presence of the forbidden state in the gap region.

In this Letter, based on the tight-binding approximation, we investigate the behavior of electron in the deformed graphene, at the critical deformation. We find the dispersion as is the generalization of the previously predicted formula in Refs. 4 and 7. We also investigate the tunneling behavior of the fermions through a gate barrier at the transition point between gapless and gapped graphene. We focus on the novel fundamental effect of the combination between the massless and massive dispersion on the tunneling through a single gate barrier.

## 2. Theory of the semi-massless fermions in graphene

Let us start with the tight-binding Hamiltonian, as given by

$$\hat{H} = -t_{ij} \sum_{<i,j>}(a_i^* b_j + H.c), \tag{1}$$

where $<i,j>$ represents the nearest neighbor sites, and $a_i$ ( $b_j$ ) is annihilation operator at sublattice A ( B) and $t_{ij}$ are the hoping energies corresponding to transferring fields among the nearest neighbor electrons. Using the Fourier transformation, $a_i = (1/\sqrt{N})\sum_k e^{i\vec{k}.\vec{R}_i} a_k$ and $b_j^* = (1/\sqrt{N})\sum_k e^{-i\vec{k}.(\vec{R}_i+\vec{\sigma}_j)} b_k^*$, where N is the number of sites of the sublattice A or B. The positions of the sublattices A and B are denoted by $\vec{R}_i$ and $\vec{R}_i + \vec{\sigma}_j$, respectively. The possible displacement vectors $\vec{\sigma}_j$ among the nearest sites for the honeycomb lattice have the three expressions $\vec{\sigma}_1 = <L_x, -L_y>$, $\vec{\sigma}_2 = <-L_x, -L_y>$ and $\vec{\sigma}_3 = <0, c'>$ as seen in Fig.1a. In the case of the undeformed graphene, $L_x = c\sqrt{3}/2$, $L_y = c/2$ and $c' = c$, where c~1.42 $A^0$ is the carbon-carbon distance of graphene. We now have the simplified formalism for the Hamiltonian in Eq.(1) of the form

$$\hat{H} = \sum_k (\omega(k) a_k^* b_k + H.c) \tag{2}$$

where $\omega(k) = -t(e^{i\vec{k}.\vec{\sigma}_1} + e^{i\vec{k}.\vec{\sigma}_2} + \eta e^{i\vec{k}.\vec{\sigma}_3})$ with t and $\eta$ being the hoping energy and asymmetric parameter due to the deformation, respectively. The Eigen energy (or the energy dispersion) for this Hamiltonian is determined by the formalism $E_k = |\omega(k)|$. Hence, we have

$$E_k = \pm \eta t \sqrt{1 + (\frac{2}{\eta})^2 \cos^2[k_x L_x] + (\frac{4}{\eta}) \cos[k_x L_x] \cos[k_y \{L_y + c'\}]}.$$

$$\tag{3}$$

The dispersion of electrons found in Eq.(3) yields the gapless graphene for $0 < \eta < 2$ and the gapped graphene for $2 < \eta$. The critical point of the transition from gapless to gapped graphene is found at $\eta = 2$. In this point, the velocity of electrons in the x-direction vanishes. By expanding the dispersion around the Dirac point $(k_D, 0) = (\frac{\pi}{L_x}, 0)$, we find that the solution of the energy dispersion of electrons can be obtained as

$$E_k \cong \pm \sqrt{(\frac{p_x^2}{2m} - \frac{p_x^4}{8(c_{eff})^2 m^3})^2 + (v_y p_y)^2}, \quad (4)$$

where $p_x = \hbar(k_x - k_D)$ and $p_y = \hbar k_y$ are the momentums in the zigzag (the x-direction) and the armchair (the y-direction) directions, respectively. Our formula in Eq.(4) is the generalization of the previously predicted formula $E_k \cong \pm \sqrt{(\frac{p_x^2}{2m})^2 + (v_y p_y)^2}$ of Refs.4 and 7. Let us next clarify the solution in Eq.4. The dispersion in the x-direction mimics the kinetic energy **T**, in the non-relativistic limit $v/c_{eff} \ll 1$, of the massive particle with mass $m = \hbar^2 / 2tL_x^2$ and with the effective speed of light $c_{eff} = (2\sqrt{3})tL_x / \hbar$, where

$$T = \sqrt{(c_{eff} p_x)^2 + (mc_{eff}^2)^2} - mc_{eff}^2 \cong \frac{p_x^2}{2m} - \frac{p_x^4}{8(c_{eff})^2 m^3} + \ldots.$$

(5)

This is to say that in the x-direction and near the Dirac point, the electronic dispersion mimics the relativistic massive kinetic energy. Therefore, our prediction shows the completed aspect of the fermion's behavior at the critical deformation. Since, in the y-direction, electrons mimic the **massless** fermions and in the x-direction, electrons mimic the **massive** fermions (see Fig1b), these particles should be called "**semi-massless fermions**".

As has been known, graphene is, in general, gapless. Recently, the possibility to open the gap in graphene has been proposed by applying the tension into a graphene sheet [8]. Applying tension in the zigzag direction can change gapless graphene to gapped graphene. Let we consider the deformed graphene system by applying tension in the **zigzag direction**. We have $L_x = (1+S)c\sqrt{3}/2$ and $L_y = c/2(1-pS)$ with the Poisson's ratio P=0.165 and S being strain. Using the hoping energy as a decaying models of $t = t_o e^{-3.37(\frac{|\vec{\sigma}_1|}{c}-1)}$ and $\eta t = t_o e^{-3.37(\frac{|\vec{\sigma}_3|}{c}-1)}$ [8], we find that the critical deformation point is found at strain of ~23%. This result leads us to have the pseudo mass of the semi-massless fermions m~$8.6\times 10^{-31}$ kg and the effective speed of light $c_{eff}$~1.4$v_F$ ($v_F$ being the Fermi velocity of the undeformed graphene). And this gives rise to the rest-mass energy $E_0 = mc_{eff}^2$ ~10.5 eV. In the case of massless fermions moving in the y-direction (or the armchair direction), we also predict $v_y$~1.09$v_F$.

**3. Tunneling of the semi-massless fermions through a single gate barrier**

We next consider the tunneling of the semi-massless fermions through a single gate barrier in the y-direction. In this direction, fermions mimic the massless particle. Therefore, we focus on the effect of the presence of the hyperbolic dispersion of the massive particle in the x-direction on the transport property of the massless fermions propagating in the y-direction. This result will compare with the original (pure) massless fermions transport. As seen in Fig.1c, the junction is biased by the voltage V and the gate potential is $V_G$(0<y<d)= -$V_G$ , vanishing elsewhere. The motion of the semi-massless fermions, with energy E and the incident angle $\theta$, is calculated using

$$[\sigma_x(-\frac{\hbar^2}{2m}\partial_x^2 + \frac{\hbar^4}{8m^2c_{eff}^2}\partial_x^4) - i\hbar v_y\sigma_y\partial_y - E_F(y) + V_G(y)]\psi(y)e^{ik_{//}x} = E\psi(y)e^{ik_{//}x}$$

(6)

where the Fermi energy $E_F(y)=E_F$, $k_{//}$ is the conservation component and $\sigma_{x,y}$ are the Pauli's spin matrices. By neglecting a small term, $-\frac{\hbar^2}{2m}\partial_x^2 \gg \frac{\hbar^4}{8m^2c_{eff}^2}\partial_x^4$ near the Dirac point, the wave motion $\psi(y)$ of electron, a solution to Eq.(6), in each region is defined as

$$\psi(y<0) = \begin{pmatrix} 1 \\ A_+ \end{pmatrix}e^{ik_yy} + r\begin{pmatrix} 1 \\ A_- \end{pmatrix}e^{-ik_yy},$$

$$\psi(0<y<d) = f\begin{pmatrix} 1 \\ B_+ \end{pmatrix}e^{ik_{yB}y} + g\begin{pmatrix} 1 \\ B_- \end{pmatrix}e^{-ik_{yB}y} \text{ and } \psi(d<y) = t\begin{pmatrix} 1 \\ A_+ \end{pmatrix}e^{ik_yy},$$

where $A_\pm = \frac{E+E_F}{(h^2k_{//}^2/2m) \mp i\hbar v_y k_y}$ and $B_\pm = \frac{E+E_F+V_G}{(h^2k_{//}^2/2m) \mp i\hbar v_y k_{yB}}$

with $k_{//} = \frac{1}{\hbar\sqrt{2}}\left(\sqrt{[(2m)^2 v_y^2 \cot^2\theta]^2 + [4(2m)^2(E+E_F)^2]} - [(2m)^2 v_y^2 \cot^2\theta]\right)^{1/2}$,

$k_y = k_{//}\cot\theta$ and $k_{yB} = \frac{1}{\hbar v_y}\sqrt{(E+E_F+V_G)^2 - (h^2k_{//}^2/2m)^2}$.

(7)

The coefficients r, f, g and t are calculated using the appropriate boundary condition at x=0 and x=d. Hence, we have the transmission coefficient as obtained by

$$t(\theta) = \frac{(A_+ - A_-)(B_+ - B_-)e^{iZ-ik_yd}}{e^{2iZ}(A_+ - B_+)(A_- - B_-) - (A_- - B_+)(A_+ - B_-)},$$

(8)

where, in the thin barrier limit $d \to 0$ and $V_G \to \infty$, $B_\pm = \pm i$, and the barrier strength $Z \sim V_G d / \hbar v_y$. The angular dependent transmission probability amplitude is defined as $T(\theta) = t^*(\theta) t(\theta)$. The dimensionless conductance of the junction is calculated using the formula $G = \int_0^{\pi/2} d\theta \cos\theta T(\theta)$.

We first consider the angular dependence of the transmission probability amplitude of the semi-fermions tunneling through a single gate barrier (see Fig.2). Here, we set E=0 as a zero biased junction. The barrier strength is also set to be of $Z \sim 0.5\pi$. The effect of combination between massless and massive fermion is investigated via the ratio parameter W=$E_F/mv_y^2$ =0.01, 0.1, 1, 10 and 1000. The **large solid line** is the plot of the transmission of the massless fermion tunneling through a single gate barrier in the original graphene, where the transmission probability amplitude has the usual expression of the form [11, 12]

$$T_{\text{original graphene}} \sim \frac{\cos^2(\theta)}{1 - \sin^2(\theta)\cos^2(Z)}.$$

The transmission of the (pure) massless fermions in undeformed graphene does not depend on the Fermi energy. This is unlike the behavior of the semi-massles fermions in the present case. Hence, it is to say that this is the effect due to the presence of the massive dispersion. In Fig.2, we find that for small W~0.1, the semi-massless fermions tunnel through the gate barrier with T~1, Klein tunneling [13], almost for all angle. Remarkably, as W is very large, the transmission decreases rapidly while the angle of incidence increases. The high transmission of the simi-massless fermions tunneling through the gate appears only the case near the normal angle $\theta = 0$, like a single channel tunneling. It is clear that this is the effect due to the presence of the massive dispersion in the x-direction to reduce T in the x-direction.

To consider the conductance of the junction (see Fig.3), we plot the conductance as a function of barrier strength Z for various W=0.01, 0.1, 1, 10 and 1000. The **large solid line** stands for the conductance of the (pure) massless fermions tunneling through a single gate barrier in the original graphene. This conventional conductance doses not depend on the Fermi energy of graphene. As a result, the zero biased conductance (E=0) of the junction is very high for $E_F << mv_y^2$ as if there was no the gate barrier at the middle. The conductance strongly oscillates while increasing W. As a result of the case where W=1000, seen in Fig.3, we may conclude that for the case of a very large W the impulse-like conductance with the maximum amplitude G=1 may be found for Z= 0, $1\pi$, $2\pi$, …., allowing the current flow for some values of the gate potential.

## 4. Summary

We have investigated the tunneling behavior of electrons through a single gate barrier in the deformed graphene. Our model may be applicable for the several honeycomb-lattice-like structures. The electronic property of electron was studied based on the tight-binding model. We obtained the carriers in graphene at the critical deformation, the transition point between gapless and gapped graphene, as the semi-massless fermions. The predicted dispersion of the carriers in graphene shows the carriers mimic the semi-massless fermions. Our predicted formula is also the generalization of the previously predicted formula. As a novel behavior, we found that, the transport property of the semi-massless-like fermions tunneling through a single gate barrier is very different from the (pure) massless fermion tunneling through a gate barrier in the original graphene structure. This is because of the effect of the presence of the combination between massless and massive dispersion in graphene at the critical deformation.

**Figure captions**

**Figure 1.** Schematic illustrations of (a) the deformed honeycomb lattice and its asymmetric electronic parameters $\vec{\sigma}_{1,2,3}$ and $t, \eta$, (b) the hyperbolic dispersion in the x-direction and the linear dispersion in the y-direction and (c) the junction with single gate barrier and the biased voltage V.

**Figure 2.** Angular dependence of the transmission of semi-fermion tunneling through a single gate barrier with various $W=E_F/mv_y^2 =0.01, 0.1, 1, 10$ and $1000$. The barrier strength is set for $Z \sim 0.5\pi$. The **large solid line** is the angular dependence of the transmission of massless fermion tunneling through a single gate barrier in the original graphene.

**Figure 3.** Plot of the conductance as a function of Z of semi-fermion tunneling through a single gate barrier with various $W=E_F/mv_y^2 =0.01, 0.1, 1, 10$ and $1000$. The **large solid line** is the plot of the conductance as a function of Z of massless fermion tunneling through a single gate barrier in the original graphene.

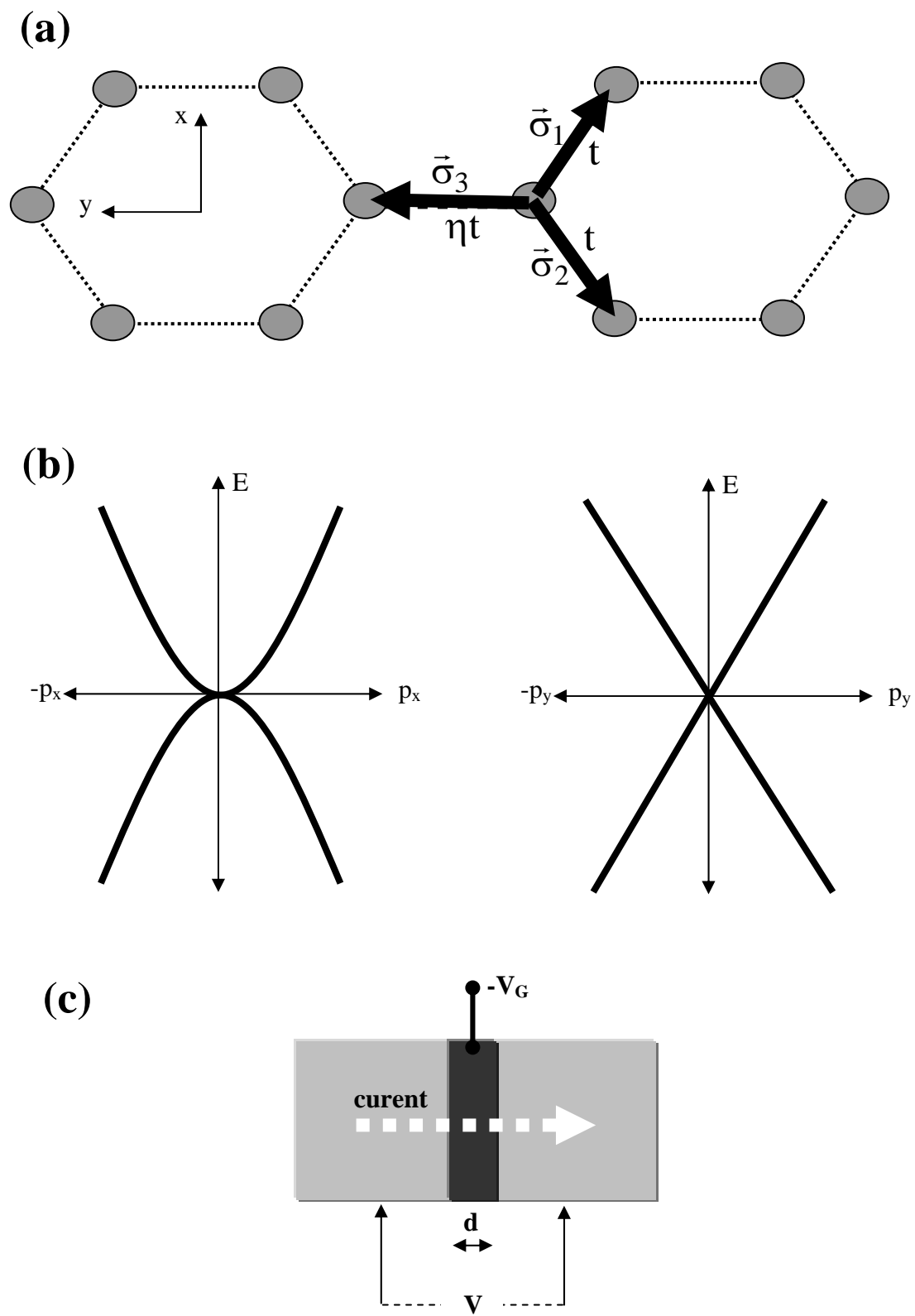

**Figure 1**

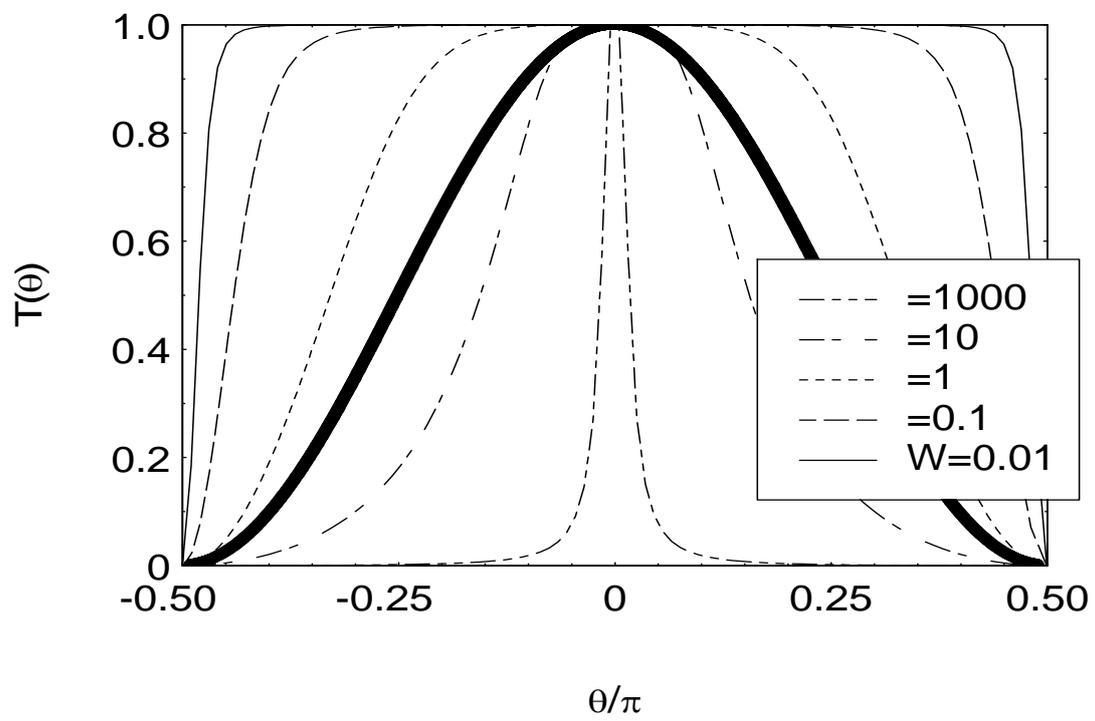

**Figure 2**

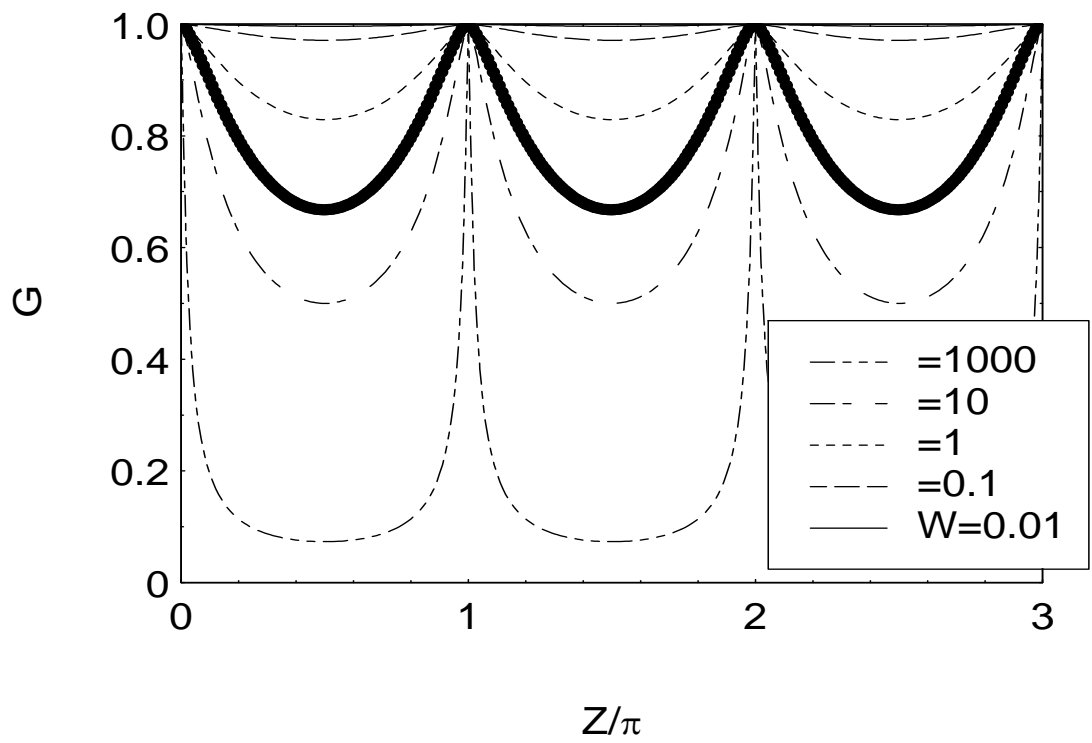

**Figure 3**